\documentclass[showpacs,preprintnumbers,amssymb]{revtex4}
\usepackage{epsfig}

\def\beq{\begin{eqnarray}}
\def\eeq{\end{eqnarray}}
\def\nnb{\nonumber}
\newcommand{\be}{\begin{equation}}
\newcommand{\ee}{\end{equation}}
\newcommand{\bea}{\begin{eqnarray}}
\newcommand{\eea}{\end{eqnarray}}
\newcommand{\ba}{\begin{array}}
\newcommand{\ea}{\end{array}}
\begin{document}
\title{$\tau^- \to \mu^- \pi^0(\eta,\eta')$ Decays
in New Physics Scenarios beyond the Standard Model}

\author{Wenjun Li}
\affiliation{ Department of Physics, Henan Normal University,
 XinXiang, Henan, 453007, P.R.China}
\author{ Yadong Yang}
\thanks{Corresponding author}
\email{yangyd@henannu.edu.cn}
\affiliation{ Department of Physics,
Henan Normal University, XinXiang, Henan, 453007, P.R.China}
\author{Xiangdan Zhang}
\affiliation{ Department of Physics, Henan Normal
University, XinXiang, Henan, 453007, P.R.China}
%\maketitle

\begin{abstract}
The semileptonic decays $\tau^- \to \mu^- M (M=\pi^0,\eta,\eta')$
could be sensitive probe for new physics scenarios with lepton
flavor violation(LFV). Motivated by the recent Belle measurement,
we investigate these decays in type III two-Higgs-doublet model
(2HDM III), R-parity violating supersymmetric models(RPV SUSY) and
flavor changing $Z'$ models with family non-universal couplings,
respectively.  In these new physics scenarios, there are LFV
couplings at tree level. Our results have shown that the decays
are very sensitive to the LFV couplings and could be enhanced to
the present experimental sensitivities. We have derived strong
bounds on relevant couplings of these models, which may be useful
for further relevant studies.

\end{abstract}


\maketitle

\noindent
\section{Introduction}
The flavor physics of fermions are among the most mysterious
fundamental problems in particle physics. In the standard
model(SM), neutrinos are exactly massless due to the absence of
the right handed chiral states ($\nu_R$) and the requirement of
$SU(2)_L$ gauge invariance and renormalizability, so that the
chirality conservation implies lepton number conservation. In the
past decade, the most exciting progress in understanding of these
issues has been the observation\cite{superK} of oscillation of
atmospheric neutrinos with very large mixing. The observation
shows that neutrinos are massive and  the lepton flavor violating
(LFV) exists in the neutral lepton sector.

In the SM supplemented with massive neutrinos, the neutrino mixing
will induce, at loop level, rare LFV processes in charge lepton
sector such as $\tau \to \mu \gamma$, $ \tau \to \mu M$(M=light
hadrons), etc. These processes are expected to be proportional to
the ratio of masses of neutrinos over the masses of the W bosons,
which is negligible small. However,  the $\tau \to \mu$ mixing
could be large in new physics models\cite{masiero}, thus the LFV
$\tau$ decays provide some sensitive probes for new physics beyond
the Standard Model.

It is interesting to note that the B factories BaBar and Belle are
also  $\tau$ factories. The $\tau$ production cross section at
BaBar and Belle is as large as $\sigma_{e^+ e^- \to \tau^- \tau^+
}=0.89nb$. The integrated luminosity up to now at Belle is about
540$fb^{-1}$, which corresponding to about $4\times 10^{8}$
$\tau^- \tau^+$ pairs. While integrated luminosity at BaBar is
about $320fb^{-1}$.  As BaBar and Belle are steadily accumulating
more data, it would be very promising to search for rare $\tau$
decays at BaBar and Belle to constrain or reveal new physics
effects.

Experimental searches have been performed for $\tau$ rare decays
associated LFV such as $\tau \to
e(\mu)\eta$~\cite{cleo,0503041ex,belle2}, $\tau \to
e(\mu)\mu^+\mu^-$~\cite{e3}, $\tau \to e(\mu)\pi^+\pi^-$~\cite{e4}
and $\tau \to e(\mu)\gamma$~\cite{e5}. The primary theoretical
studies are focused on $\mu$ and $\tau$ radiative decays  and
their decays to three charged leptons in different
models~\cite{omodels,sking}. There are also many studies on decays
$\tau \to \ell M$. Ilakovac et al., have studied  $\tau \to e(\mu)
M$ decays in models with heavy Dirac or Majorana neutrinos and
found $\tau \to e\phi(\rho^0,\pi^0)$ with branching ratios of
order $10^{-6}$~\cite{ilak}. Sher has investigated  the decays
$\tau \to \mu \eta$ in the large tan $\beta$ region in seesaw
MSSM~\cite{msher}. In MSSM model, these decays are analyzed by
Fukuyama \textit{et al} ~\cite{Fuku}. Brignole and
Rossi\cite{rossi} has presented a comprehensive theoretical study
on these decays
 recently  in a general unconstrained MSSM.
 A model independent study has been preformed by
Black $\textit{et
 al}$ \cite{han} to bound new physics scales.

In order to get information on LFV couplings, an important
observable, the anomalous magnetic moment of the muon $(g-2)_\mu$,
should  be considered additionally. Several works constraining LFV
processes from previous estimations on $a_\mu$ have been carried
out in Refs\cite{a-mu,0103064}.

The most recent experimental search for semileptonic LFV $\tau$
transitions has performed by Belle~\cite{0503041ex} using
\textit{only} $153.8 fb^{-1}$  data
 \beq
&&\mathcal{B}(\tau \to \mu \pi^0) < 4.1  \times 10^{-7},~~90\%~CL\nnb \\
&&\mathcal{B}(\tau \to \mu \eta) < 1.5 \times 10^{-7},~~90\% ~CL \nnb \\
&&\mathcal{B}(\tau \to \mu \eta') < 4.7 \times 10^{-7},~~90\% ~CL.
 \eeq
These results are already 10 to 64 times more restrictive than
previous CLEO limits\cite{cleo}. Using the theoretical formula
derived by Sher\cite{msher} in a seesaw MSSM, the Belle new
results have improved the constraints of the allowed parameter
space for $m_A -\tan\beta$ in MSSM. It would be very worthy to
study these decays in other new physics models to derive bounds on
relevant parameters. In this paper, we will study these decays in
three new physics scenarios, namely,
\begin{itemize}

  \item the 2HDM III\cite{2hd,soni,mo3}, where
   FCNC and LFV could arise at the tree level in the Yukawa sector when
   the up-type quarks and down quarks are allowed simultaneously to couple
   more than one scalar doublet,

   \item SUSY theories with R-parity
   broken~\cite{RPVm1,RPVm2,report}, in which the R parity odd interactions can
   violate the lepton and baryon numbers as well as couple the
    different generations or flavors of leptons and quarks,

   \item flavor changing $Z^{\prime}$ models\cite{Z1,Z2,Z3} with
    family non-universal couplings.

\end{itemize}

We have shown that the LFV semileptonic $\tau$ decays could be
enhanced to the present B factories sensitives in the above three
scenarios, thus we have derived bounds on the LFV couplings in the
models, which are tighter than exiting ones in the literature.

In next section, we present calculations of the decays and bounds
on the LFV couplings in the aforementioned three scenarios.
Finally in Sec.III, we give our conclusions.

\section{Model calculations}

\subsection{Hadronic matrices of local operators }
Before detail model calculations of $\tau\to \mu M (M=\pi^0, \eta,
\eta')$, we would specify hadronic matrices elements which are
inputs for calculating these decay amplitudes.

At first, we need
 \beq
 \langle \pi^0 (p)|q\gamma^{\mu}\gamma_5
 q|0\rangle=-\frac{i}{\sqrt{2}}f_{\pi}p_{\mu}
 \eeq
with $f_{\pi}=130\pm5$ MeV,  and the so-called chiral condensation
matrix
  \beq
 \langle \pi^0 (p)|q\gamma_5
 q|0\rangle=-\frac{i}{\sqrt{2}}f_{\pi}\frac{m_{\pi}^2}{2m_q}
 \eeq
where q=u or d. We will see that the $1/2m_q$ factor will cancel
the corresponding quark mass coupling of the weak scalar
interaction operators, and thus enhances scalar interaction
contributions.

As for $\eta$ and $\eta'$, the situation is much more complicated
than $\pi^0$. The relevant matrices are defined by
 \beq
 \langle M(p)|\bar{q} \gamma ^\mu
\gamma_5 q|0 \rangle = -\frac{i}{\sqrt{2}}f^q_M p^\mu ,
 &&
~~~~~2m_q\langle
M(p)|\bar{q}\gamma_5 q|0 \rangle = -\frac{i}{\sqrt{2}}h^q_{M}, \nonumber \\
 \langle M(p)|\bar{s}\gamma^\mu\gamma_5 s|0 \rangle =
-if^s_M  p^\mu ,
 &&
 ~~~~~2m_s\langle M(p)|\bar{s}
\gamma_5 s|0 \rangle = -i h^s_{M}, \label{hi}
  \eeq
 where $M=\eta$ or $\eta'$. These equations define eight
 non-perturbative parameters, which however are not all
 independent. They could be related to fewer independent non-perturbative parameters
 by  $\eta-\eta'$ mixing scheme.
  In this paper, we will take the
Feldmann-Kroll-Stech (FKS) mixing scheme\cite{9812269}. In FKS
mixing scheme the parton Fock state decomposition can be expressed
as
 \beq
\left( \ba{c} |\eta\rangle\\
|\eta'\rangle \ea \right) =
\left( \ba{c c} \cos\phi & -\sin\phi\\
\sin \phi & \cos \phi \ea\right) \left( \ba{c}
|\eta_q\rangle\\|\eta_s\rangle \ea \right),
\eeq
 where $\phi$ is
the mixing angle,  $|\eta_q \rangle =(|u \bar{u}\rangle + |d
\bar{d}\rangle)/\sqrt{2}$ and $|\eta_s\rangle =|s \bar{s}\rangle$.
The four parameters $f^i_M$ are therefore related by
 \beq
&f^q_\eta = f_q \cos \phi, &~~~~ f^s_\eta = -f_s \sin \phi,
\nonumber \\
& f^q_{\eta'} = f_q \sin\phi, &~~~~  f^s_{\eta'} = f_s\cos\phi,
 \eeq
and an analogous set of equations for the $h^i_M$
 \beq
&h^q_\eta = h_q \cos \phi, &~~~~ h^s_\eta = -h_s \sin \phi,
\nonumber \\
& h^q_{\eta'} = h_q \sin\phi, &~~~~  h^s_{\eta'} = h_s\cos\phi.
 \eeq
Moreover the $h_q$ and $h_s$ could be related to $f_q$, $f_s$ and
mixing angle $\phi$ \cite{martin}
 \beq
 h_q &=&f_q (m^2_{\eta} \cos^2 \phi +m^2_{\eta'} \sin^2 \phi)-\sqrt{2} f_s
          (m^2_{\eta'}-m^2_{\eta})\sin\phi \cos\phi,\nonumber \\
 h_s &=&f_s (m^2_{\eta'} \cos^2 \phi +m^2_{\eta} \sin^2 \phi)
         -\frac{f_q}{\sqrt{2}}(m^2_{\eta'}-m^2_{\eta})\sin\phi
         \cos\phi.
 \eeq
The three remaining parameters $f_q$, $f_s$ and $\phi$  in FKS
 scheme have been constrained from the available experimental data with
results\cite{9812269}
 \beq
 f_q =(1.07\pm0.02)f_{\pi}, ~~f_s =(1.34\pm 0.06)f_{\pi},
 ~~\phi =39.3^{\circ}\pm 1.0^{\circ}.
 \eeq
From these parameters and $f_{\pi}$, Beneke and
Neubert\cite{martin} have derived
 \beq \label{hinputs}
  f^q_{\eta} =108\pm 3MeV,~~   f^s_{\eta} =-111\pm 6MeV,~~
  h^q_{\eta} =0.001\pm 0.003GeV^3,~~  h^s_{\eta} =-0.055\pm 0.003GeV^3,\nonumber\\
  f^q_{\eta'} =89\pm 3MeV,~~~    f^s_{\eta'} =136\pm 6MeV,~~~
  h^q_{\eta'} =0.001\pm 0.002GeV^3,~~~  h^s_{\eta} =0.068\pm
  0.005GeV^3.
 \eeq
 It should noted that $h^q_{\eta}$ and $h^q_{\eta'}$ are poorly
 determined. In our numerical calculations, we take these hadronic
 parameters with $1\sigma$ variant to display their uncertainty
 effects on our bounding on LFV couplings.
 Now we are ready to calculate the decays in the
aforementioned three new physics scenarios.

\subsection{$\tau\to \mu M$ decays in  2HDM III}

Two Higgs Doublet Model(2HDM) is the popular and the most simplest
extension of the SM with a scalar sector made of two instead of
one complex scalar doublets. In order to build a 2HDM without FCNC
at tree level, it is achieved by requiring either that
\textit{u}-type and \textit{d}-type quarks couple to the same
doublet(2HDMI) or  that \textit{u}-type quarks couple to one
scalar doublet and \textit{d}-type quarks to the other(2HDMII).
The case in which scalar FCNC not forbidden is  dubbed 2HDMIII,
where the Higgs doublets could couple to both the \textit{u}- and
\textit{d } type quarks  at the same time\cite{soni,mo3}.

Generally one can write  Yukawa Lagrangian of
2HDMIII~\cite{soni,mo3,prd3484}
 \beq {\cal L}_{Y}= \eta^{U}_{ij}
\bar Q_{i,L} \tilde H_1 U_{j,R} + \eta^D_{ij} \bar Q_{i,L} H_1 D_{j,R} + \xi^{U}_{ij}
\bar Q_{i,L}\tilde H_2 U_{j,R} +\xi^D_{ij}\bar Q_{i,L} H_2
D_{j,R} \,+\, h.c., \label{lyukmod3}
 \eeq
 where $H_i(i=1,2)$ are the two Higgs doublets. $Q_{i,L}$ is the left-handed
 fermion doublet,  $U_{j,R}$ and
$D_{j,R}$ are the right-handed singlets, respectively. These
$Q_{i,L}, U_{j,R}$ and $D_{j,R}$ are weak eigenstates, which can
be rotated into mass eigenstates, while $\eta^{U,D}$ and
$\xi^{U,D}$ are the non-diagonal matrices of the Yukawa couplings.

For convenience one can express $H_1$ and $H_2$ in a suitable basis such that only
the $\eta_{ij}^{U,D}$ couplings generate the fermion masses, i.e.,
 \be
 \langle H_1\rangle=\left(
\ba{c}
0\\
\frac{v}{\sqrt{2}}\ea \right), \,\,\,\, \,\,\,\, \langle H_2\rangle=0.
 \ee
 The two doublets in this basis  have the form
 \be
 \label{base}
 H_1=\frac{1}{\sqrt{2}}\left[\left(\ba{c} 0 \\
v+\phi^0_1 \ea\right)+ \left(\ba{c} \sqrt{2}\, G^+\\
i G^0\ea\right)\right], \,\,\,\,\,\,\,\,
 H_2=\frac{1}{\sqrt{2}}
 \left(\ba{c}\sqrt{2}\,H^+\\ \phi^0_2+i A^0\ea\right),
 \ee
where $G^{0,\pm}$ are the Goldstone bosons, $H^{\pm}$ and $A^0$
are the physical charged-Higgs boson and CP-odd neutral Higgs
boson respectively. The advantage of choosing the basis is  the
first doublet $H_1$ corresponding to the scalar doublet of the SM
while the new Higgs fields arising from the second doublet $H_2$.
So $H^0_2$ does not couple  to  gauge bosons of the form $H^0_2ZZ$
and $H^0_2W^+W^-$.

In Eq.(\ref{base}), $\phi_1^0$ and $\phi^0_2$ are not the neutral
mass eigenstates but linear combinations of the CP-even neutral
Higgs boson mass eigenstates $H^0$ and $h^0$ \beq
\label{masseigen} H^0 & = & \phi_1^0 \cos\alpha +
\phi^0_2\sin\alpha ,\\ h^0 & = & -\phi^0_1\sin\alpha + \phi^0_2
\cos\alpha   ,
 \eeq
 where $\alpha$ is the mixing angle.  In the case of
$\alpha=0$, ($\phi^0_1$, $\phi^0_2$) coincide with the mass
eigenstates of $H^0$ and $h^0$.

After diagonalizing the mass matrix of the fermion fields, the
Yukawa Lagrangian becomes
 \be {\cal L}_{YC} = \hat\eta^{U}_{ij}
\bar Q_{i,L} \tilde H_1 U_{j,R} + \hat\eta^D_{ij} \bar Q_{i,L} H_1 D_{j,R}
+ \hat\xi^{U}_{ij} \bar Q_{i,L}\tilde H_2 U_{j,R}
 +\hat\xi^D_{ij}\bar Q_{i,L} H_2 D_{j,R} \,+\, h.c. ,\label{lyukmass}
 \ee
where $Q_{i,L}$, $U_{j,R}$, and $D_{j,R}$ now denote the fermion
mass eigenstates and
 \beq
\hat\eta^{U,D}&=&(V_L^{U,D})^{-1}\cdot \eta^{U,D} \cdot
V_R^{U,D}=\frac{\sqrt{2}}{v}M^{U,D}(M^{U,D}_{ij}=\delta_{ij}m_j^{U,D}),\label{diag}\\
\hat\xi^{U,D}&=&(V_L^{U,D})^{-1}\cdot \xi^{U,D} \cdot V_R^{U,D} \label{neutral}.
 \eeq
 In Eq.(\ref{neutral}), $V_{L,R}^{U,D}$ are
the rotation matrices acting on  up and down-type quarks, with
left and  right chiralities
 respectively. Thus $V_{CKM}=(V_L^U)^{\dag}V_L^D$ is the
usual Cabibbo-Kobayashi-Maskawa (CKM) matrix. We can see from
Eq.(\ref{lyukmass}) that the matrices $\hat\xi^{U,D}$, as defined
by Eq.(\ref{neutral}), allow scalar-mediated FCNC. That is, in the
quark mass basis only the matrices $\hat\eta^{U,D}$ of
Eq.(\ref{diag}) are diagonal, but the matrices $\hat\xi^{U,D}$ are
in general not diagonal. The FCNC part of the Yukawa Lagrangian is
 \beq
 {\cal L}_{Y,FCNC} &=& -\frac{H^0\sin\alpha+h^0\cos\alpha}{\sqrt{2}}\left\{
 \bar{U}\biggl[\hat\xi^{U}\frac{1}{2}(1+\gamma_5)
 +\hat\xi^{U\dagger}\frac{1}{2}(1-\gamma_5)\biggl]U
 +\bar{D}\biggl[\hat\xi^{D}\frac{1}{2}(1+\gamma_5)
 +\hat\xi^{D\dagger}\frac{1}{2}(1-\gamma_5)\biggl]D
 \right\} \nonumber \\&&+ \frac{iA^0}{\sqrt{2}}\left\{
 \bar{U}\biggl[\hat\xi^{U}\frac{1}{2}(1+\gamma_5)
 -\hat\xi^{U\dagger}\frac{1}{2}(1-\gamma_5)\biggl]U
-\bar{D}\biggl[\hat\xi^{D}\frac{1}{2}(1+\gamma_5)
 -\hat\xi^{D\dagger}\frac{1}{2}(1-\gamma_5)\biggl]D
 \right\}.\label{lyukfc}
 \eeq
The corresponding  Feynman rules from Eq.(\ref{lyukfc}) can be
found in Refs.~\cite{soni,9811235}.

Because the definition of  $\xi^{U,D}_{ij}$ couplings is
arbitrary, we can take the rotated couplings as the original ones
and shall write $\xi^{U,D}$ in stead of $\hat{\xi}^{U,D}$
hereafter.

In this paper, we use the Cheng-Sher ansatz~\cite{prd3484}
 \be \xi^{U,D}_{ij}=\lambda_{ij}
\,\frac{\sqrt{m_i m_j}}{v} \label{sher}
 \ee
 which ensures that the FCNC within the first two
generations are naturally suppressed by small fermions masses.
This ansatz suggests that LFV couplings involving the electron are
naturally suppressed, while LFV transition involving muon and tau
are much less suppressed and may generate sizeable effects.

The decay amplitudes are given by
\begin{eqnarray}\label{2hdm3:mipi}
{\cal A}(\tau^- \to \mu^- \pi^0)&=& \frac{G_F}{4}
\sqrt{m_\mu m_\tau}f_{\pi^0}m^2_{\pi^0}
 \left\{E\lambda_{\tau\mu}(\bar{\mu}\tau)_{S+P}
+F\lambda^*_{\tau\mu}(\bar{\mu}\tau)_{S-P}\right\},  \\
\label{2hdm3:mieta}{\cal A}(\tau^- \to \mu^- \eta) &=&
\frac{G_F}{2\sqrt{2}}\sqrt{m_\mu m_\tau} \left\{\biggl[h^q_{\eta}
J_q + h^s_{\eta} J_s\biggl]\lambda_{\tau\mu}(\bar{\mu}\tau)_{S+P}
 +\biggl[h^q_{\eta} K_q + h^s_{\eta} K_s\biggl]
\lambda^*_{\tau\mu}(\bar{\mu}\tau)_{S-P}\right\},  \\
\label{2hdm3:mietap} {\cal A}(\tau^- \to \mu^- \eta')
&=&\frac{G_F}{2\sqrt{2}}\sqrt{m_\mu m_\tau}
\left\{\biggl[h^q_{\eta'} J_q +h^s_{\eta'}
J_s\biggl]\lambda_{\tau\mu}(\bar{\mu}\tau)_{S+P}
+\biggl[h^q_{\eta'} K_q+ h^s_{\eta'}
K_s\biggl]\lambda^*_{\tau\mu}(\bar{\mu}\tau)_{S-P}\right\},
\end{eqnarray}
where   $h^{i}_{M}$ are defined by Eq.\ref{hi}  and the auxiliary
functions are
\begin{eqnarray}
E&=&\frac{1}{m^2_{A^0}}\biggl(Re\lambda_{uu}+Re\lambda_{dd}\biggl)
+i\biggl(\frac{\sin^2\alpha}{m^2_{H^0}}
+\frac{\cos^2\alpha}{m^2_{h^0}}\biggl)
\biggl(Im\lambda_{uu}-Im\lambda_{dd}\biggl), \nonumber \\
F&=&-\frac{1}{m^2_{A^0}}\biggl(Re\lambda_{uu}+Re\lambda_{dd}\biggl)
+i\biggl(\frac{\sin^2\alpha}{m^2_{H^0}}
+\frac{\cos^2\alpha}{m^2_{h^0}}\biggl)\biggl(Im\lambda_{uu}-Im\lambda_{dd}\biggl), \\
J_q&=&\frac{1}{m_u+m_d}\left[\frac{1}{m^2_{A^0}}
\biggl(m_u Re\lambda_{uu}-m_d Re\lambda_{dd}\biggl)
+i\biggl(\frac{\sin^2\alpha}{m^2_{H^0}}
+\frac{\cos^2\alpha}{m^2_{h^0}}\biggl)
\biggl(m_u Im\lambda_{uu}+m_dIm\lambda_{dd}\biggl)\right],\nonumber \\
J_s&=&\left[-\frac{1}{m^2_{A^0}}Re\lambda_{ss}+i\biggl(\frac{\sin^2\alpha}{m^2_{H^0}}
+\frac{\cos^2\alpha}{m^2_{h^0}}\biggl)Im\lambda_{ss}\right],\nonumber \\
K_q&=&\frac{1}{m_u+m_d}\left[ -\frac{1}{m^2_{A^0}}
\biggl(m_u Re\lambda_{uu}-m_d Re\lambda_{dd}\biggl)
+i\biggl(\frac{\sin^2\alpha}{m^2_{H^0}}
+\frac{\cos^2\alpha}{m^2_{h^0}}\biggl)
\biggl(m_u Im\lambda_{uu}+m_d Im\lambda_{dd}\biggl)\right],\nonumber \\
K_s&=&\left[ \frac{1}{m^2_{A^0}}Re\lambda_{ss}
+i\biggl(\frac{\sin^2\alpha}{m^2_{H^0}}
+\frac{\cos^2\alpha}{m^2_{h^0}}\biggl)Im\lambda_{ss}\right].
\nonumber
\end{eqnarray}
From Eq.\ref{2hdm3:mipi}-\ref{2hdm3:mietap}, we can see that there are two types operators
 $(\bar{\mu}\tau)_{S\pm P}$ contribute to decays at
tree-level. And it should be noted that the small quark mass
factor in the neutral Higgs- quark couplings are cancelled by the
one in   $\langle P^0|\bar{q}\gamma_5q|0 \rangle$.
\begin{table}[htb]\label{ql}
\caption{Constraints on the $\lambda_{\tau\mu}$ from $\tau^- \to
\mu \pi^0,  \mu \eta^{(')} $ decays}
\begin{center}
\begin{tabular}{c c c}
\hline \hline Decay modes&~~~~ Bounds on $\lambda_{\tau\mu}$& Previous Bounds\\
\hline
 $\tau \to \mu \pi^0$&$\leq 63.62$&\ \ \
  $\lambda_{\tau\mu}\sim O(1)$~\cite{prd3484}, \ \ \ $\lambda_{\tau\mu}
  \sim O(10)-O(10^2)$~\cite{0208117}
\\
 $\tau \to \mu \eta$&$\leq 2.76$ &  $\lambda_{\tau\mu} \sim O(10^2)-O(10^3)$~\cite{0501162} \\
 $\tau \to \mu \eta'$&$\leq 4.48$& $|\lambda_{\mu\mu}|
 =|\lambda_{\tau\tau}|=|\lambda_{\mu\tau}|=|\lambda_{e\mu}|
 =10$~\cite{0502087}\\ \hline \hline
\end{tabular}
\end{center}
\end{table}

In this work we choose $\xi^{U,D}$ to be complex for the sake of
simplicity, so that besides Higgs boson masses, only
$\lambda_{uu}, \lambda_{dd}$ and $\lambda_{ss}$ in the quark
sector and $\lambda_{\tau\mu}$ in the lepton sector are parameters
relevant to the semileptonic $\tau$ decays. Totally there are
seven parameters in the amplitudes:
$\lambda_{qq}(q=u,d,s),\lambda_{\tau\mu}$, their phases $\theta$,
the masses of neutral Higgs $m_{h^0}, m_{H^0}, m_{A^0}$ and the
mixing angle $\alpha$. Taking into account of  constraints on
parameters from experimental data and theoretical limits, the
values of parameters can be taken as~\cite{dyb,csh-mk,0208117}
 \beq
\theta=\pi/4,~~|\lambda_{uu}|=150,~~|\lambda_{dd}|=120,\,|\lambda_{ss}|=100,\
 m_{h^0}=115GeV,\ m_{A^0}=120GeV,~~m_{H^0}=150GeV,~~\alpha=\pi/4.
\eeq
 where the higgs masses satisfy the relation
  $115GeV \leq m_{h^0} < m_{A^0}<m_{H^0}\leq 200GeV$\cite{dyb,csh-mk,0208117}.

\begin{figure}\label{2HDM3}
\scalebox{0.5}{\epsfig{file=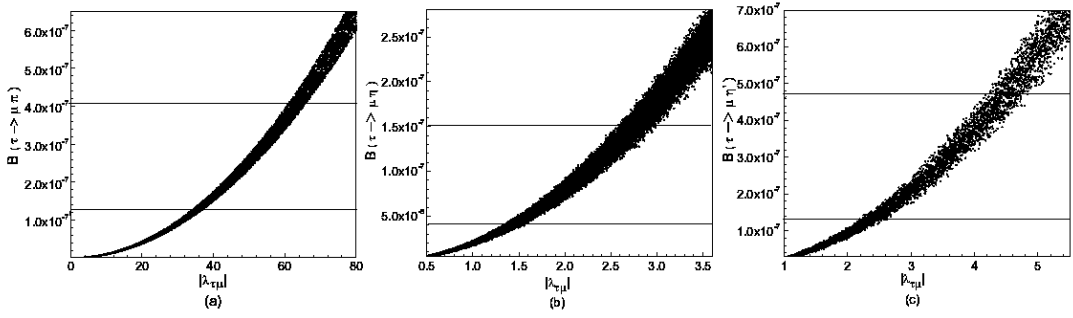}}  \caption{ The branching
ratios as functions of $|\lambda_{\tau\mu}|$ for fixed
$m_{H^0}=150 GeV, m_{A^0} =120 GeV, m_{h^0}=115 Gev,\alpha=\pi/4$,
 the phase of $\lambda_{\tau\mu},\theta=\pi/4$ in the 2HDM III.
(a)  for $\tau^- \to \mu^- \pi^0$ decay, (b) for $\tau^- \to \mu^-
\eta$ decay, and (c)  for $\tau^- \to \mu^- \eta'$ decay. The
upper horizontal lines denote Belle upper bounds. The lower
horizontal lines denote the possible upper bounds obtained naively
by scaling Belle upper bounds with $540fb^{-1}$ integrated
luminosity. The dotted bands correspond to the effects of hadronic
inputs uncertainties.}
\end{figure}
In the literature, the effects of  $\lambda_{ij}$ have been
studied under different phenomenological
considerations~\cite{soni,9811235,prd3484,dyb,csh-mk,0208117,
0501162,0502087,0010149}. For comparison we list  bounds on the
$\lambda_{\tau\mu}$ parameter of 2HDM III in Table.I. In
Ref.~\cite{prd3484}, it is suggested $\lambda_{\tau\mu}\sim O(1)$.
However, based on limits from $(g-2)_\mu$ results~\cite{0208117}
and tau decay~\cite{0501162}, $\lambda_{\tau\mu}$ could have much
larger value $\sim O(10-10^3)$. The branching ratios for $\tau^-
\to \mu^- \pi^0(\eta,\eta')$ versus $|\lambda_{\tau\mu}|$ in 2HDM
III are presented in Fig.1, where Fig.1.(a), Fig.1.(b) and
Fig.1.(c) are the results for $\tau^-\to\mu^-\pi^0$,
$\tau^-\to\mu^-\eta$ and $\tau^-\to\mu^-\eta'$ decays,
respectively.  The dotted bands show our theoretical uncertainties
due to the variants of hadronic inputs as in Eq.(\ref{hinputs}).
The hadronic uncertainties  in $\mathcal{B}(\tau\to \mu \pi^0)$
arise from $f_{\pi}$ and are very small. Although
$\mathcal{B}(\tau\to \mu \eta,~ \mu \eta^{\prime)})$ involve the
poorly known parameters $h^q_{\eta}$ and $h^q_{\eta'}$, the two
decays are dominated by $h^s_{\eta}$ and $h^s_{\eta'}$,
respectively. So the hadronic uncertainties are moderate. The
upper horizonal lines denote current \textit{Belle } upper limits.
As shown in Fig.1, the branching ratios, as functions of
$|\lambda_{\tau\mu}|$, rise rapidly with the increase of
$|\lambda_{\tau\mu}|$. We find that  $\tau^- \to \mu^-
\eta(\eta')$ decays are more sensitive to $|\lambda_{\tau\mu}|$
than $\tau^-\to\mu^- \pi^0$ decay. From the Belle
measurement\cite{0503041ex}, we get bounds on the LFV coupling
$|\lambda_{\tau\mu}|$ which are listed in Table.I for  hadronic
inputs with central values defaulted.   $\tau^- \to \mu^-
\eta,\eta'$ decays give upper bounds on the strength of
$|\lambda_{\tau\mu}|$ at order of $ \mathcal{O}(1)$, while $\tau^-
\to \mu^- \pi^0$ gives a looser bound. As shown in Table.I,
comparing to the former constraints from other processes, we have
obtained much more stringent bound for $|\lambda_{\tau\mu}|$ from
$\tau\to \mu\eta^{(\prime)}$. Through calculations, we find that
the branching ratios of $\tau^- \to \mu^- \pi$
 can be as low as $2.0\times 10^{-9}$ when $\lambda_{\tau\mu} \simeq 4.5$.
 It should be noted  that the current Belle bounds are based on
 \textit{only } $153.8fb^{-1}$  data. Up to now, Belle has
 accumulated about $540fb^{-1}$ data already.
 To show the potential of bounding on LFV couplings with the full data at Belle,
 we scale the currents upper bounds \textit{ naively } by a factor
  $153.8fb^{-1}/540fb^{-1}$.
As a benchmark, the potential are presented  by lower horizontal
lines in Fig.1, which would give bounds  few times more
restrictive  than these from current Belle upper limits.

 To conclude this subsection, we have shown that, at the similar experimental
  sensitivity to the three decay modes, searching
for $\tau\to\mu \eta'$ would put more tighten constraints on the
Higgs couplings than these from $\tau\to\mu\pi^0$ decays.

\subsection{ $\tau\to \mu M$ in RPV SUSY model}

The Supersymmetry model with explicit R-Parity breaking provides a
simple framework for neutrino masses and mixing angles in
agreement with the experimental data\cite{allan, report}.  The R
parity quantum number is defined by\cite{RPVm1}
 \beq R=(-1)^{3B+L+2S},
 \eeq
 where B is  the
baryon number, L the lepton number and S the spin, respectively.

Apparently, the lepton  and/or baryon number violation could lead
to R-Parity violation. The explicit R-Parity breaking would
introduce renormalizable bilinear higgsino-lepton field mixings
and trilinear Yukawa couplings between fermions and their
super-partners \cite{report}
 \beq
 W_{RPV}&=& \sum_i \mu_iL_iH_u
+\sum_{i,j,k}\biggl(\frac{1}{2}\lambda_{ijk}L_iL_jE^c_k
+\lambda'_{ijk}L_iQ_jD^c_k+\lambda^{''}_{ijk}U^c_iD^c_jD^c_k\biggl),
\\
L_{RPV}&=&\sum_i\mu_i(\bar{\nu}_{iR}\tilde{H}^{0c}_{uL} -\bar{e}_{iR}\tilde{H}^{
+c}_{uL})+\sum_{i,j,k}\left\{\frac{1}{2}\lambda_{ijk}
\biggl[\tilde{\nu}_{iL}\bar{e}_{kR}e_{jL}+\tilde{e}_{jL}\bar{e}_{kR}\nu_{iL}
+\tilde{e}^\star_{kR}\bar{\nu}^c_{iR}e_{jL}-(i\rightarrow
j)\biggl]\right.\nnb
\\ &+&\left. \lambda'_{ijk}\biggl[\tilde{\nu}_{iL}\bar{d}_{kR}d_{jL}
+\tilde{d}_{jL}\bar{d}_{kR}\nu_{iL}+\tilde{d}^\star_{kR}\bar{\nu}^c_{iR}d_{jL}
-\tilde{e}_{iL}\bar{d}_{kR}u_{jL}-\tilde{u}_{jL}\bar{d}_{kR}
e_{iL}-\tilde{d}^\star_{kR}\bar{e}^c_{iR}u_{jL}\biggl]\right.\nnb \\
&+&\left.\frac{1}{2}\lambda^{''}_{ijk}
\epsilon_{\alpha\beta\gamma}\biggl[\tilde{u} ^\star_{i\alpha
R}\bar{d}_{j\beta R}d^c_{k\gamma L} +\tilde{d}^\star_{j\beta
R}\bar{u}_{i\alpha R}d^c_{k\gamma L}+\tilde{d}^\star _{k\gamma
R}\bar{u}_{i\alpha R}d^c_{j\gamma L}\biggl]\right\},
 \eeq
 where the indices $i,j,k(=1,2,3)$ label  quark and lepton
generations. $L_i$ and $Q_i$ are the SU(2)-doublet lepton and
quark superfields. $U^c_i,D^c_i,E^c_i$ are the singlet
superfields. $\lambda_{ijk}$ is antisymmetric under the
interchange of the first two SU(2) indices, while
$\lambda^{''}_{ijk}$ is antisymmetric under the interchange of the
last two. For the detailed reviews on R-parity violation SUSY
model could be found in~\cite{book,9608415,9810232,9906209}. The
phenomenological constraints on R-parity violation couplings have
been studied extensively in various processes  with different
points of view~\cite{collider,rpvs,rpv,pko,Barbier,chemtob}. For
comparison, we list the constraints on $\lambda,\lambda'$
couplings relevant to $\tau \to \mu M$ in Table II, where the
upper limits on $\lambda$ and $\lambda'$ are generally
$\mathcal{O}(10^{-2})$ expect constraints from $dd\to uu e^-e^-$
and $K \to \pi \nu\nu$.

In the RPV SUSY model, the amplitudes of these decays are
calculated to be
\begin{eqnarray}\label{rpv:mp}
{\cal A}(\tau^- \to \mu^- \pi^0)&=&
\frac{1}{8\sqrt{2}}f_{\pi^0}\left\{-\frac{m^2_{\pi^0}}{2m_d}
 L_1(\bar{\mu}\tau)_{S-P} + L_2
 p^\mu_{\pi^0}(\bar{\mu}\tau)_{V+A} + L_3 p^\mu_{\pi^0}
(\bar{\mu}\tau)_{V-A} \right\} ,\\
\label{rpv:me}{\cal A}(\tau^- \to \mu^- \eta)&=&\frac{1}{8}
\left\{\frac{ p^\mu_\eta f_{\eta}^q}{2} L_2 (\bar{\mu}\tau)_{V+A}
 -\left[ p^\mu_\eta f_{\eta}^s
  L_4 +\frac{p^\mu_\eta f_{\eta}^q}{2} L_3\right]
(\bar{\mu}\tau)_{V-A}\right.\nonumber \\
&&\left.+\left[ \frac{h_{\eta}^q}{2(m_u+m_d)} L_1+
\frac{h_{\eta}^s}{2m_s} L_5\right](\bar{\mu}\tau)_{S-P}\right\},\\
\label{rpv:mep}{\cal A}(\tau^- \to \mu^- \eta^{'})&=&
 \frac{1}{8}\left\{ \frac{p^\mu_{\eta^{'}}
   f_{\eta^{'}}^q}{2} L_2 (\bar{\mu}\tau)_{V+A} -\left[p^\mu_{\eta^{'}} f_{\eta^{'}}^s
  L_4 +  \frac{p^\mu_{\eta^{'}} f_{\eta^{'}}^q}{2}
L_3\right] (\bar{\mu}\tau)_{V-A}\right.\nonumber \\
&&\left.+\left[ \frac{h_{\eta^{'}}^q}{2(m_u+m_d)} L_1+
 \frac{h_{\eta^{'}}^s}{2m_s} L_5\right](\bar{\mu}\tau)_{S-P}\right\},\eeq
with \beq
L_1=\frac{\lambda'_{i11}\lambda_{i23}}{m^2_{\tilde{\nu}}}, \ \ \ \
\ \ L_2=\frac{\lambda'_{21k}\lambda^{'*}_{31k}}{m^2_{\tilde{d}}},\
\ \ \ \ \
L_3=\frac{\lambda'_{3j1}\lambda^{'*}_{2j1}}{m^2_{\tilde{u}}},\ \ \
\ \ \
L_4=\frac{\lambda'_{3j2}\lambda^{'*}_{2j2}}{m^2_{\tilde{u}}},\ \ \
\ \ \ L_5=\frac{\lambda'_{i22}\lambda_{i23}}{m^2_{\tilde{\nu}}}.\
\ \ \
\end{eqnarray}
%%%%%%FIG2%%%%%%
\begin{figure} \label{RPVpi}
\scalebox{0.9}{\includegraphics[width=15cm] {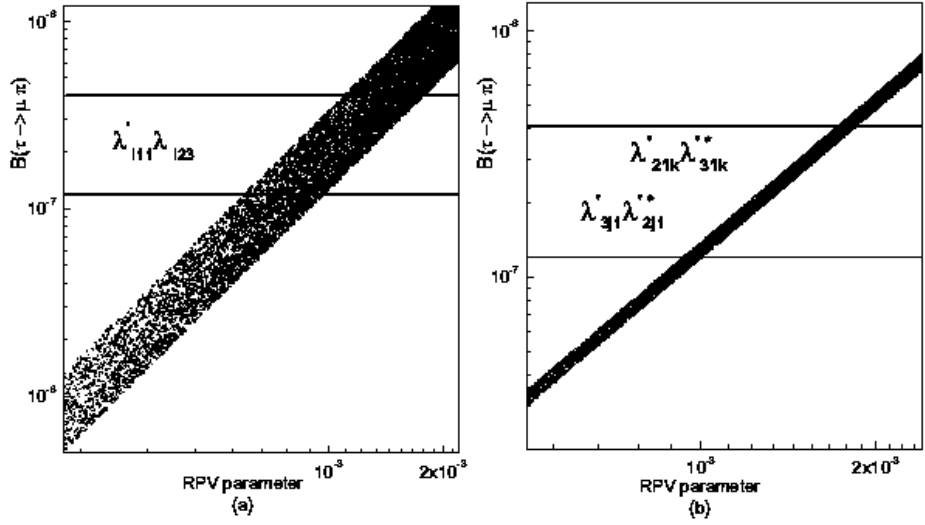}}
\caption{$\mathcal{B}(\tau\to\mu\pi)$ as function of R-parity
violation couplings.  (a)  for $|\lambda'_{i11}\lambda_{i23}|$,
(b)  for $|\lambda'_{21k}\lambda^{'*}_{31k}|$ and
$|\lambda'_{3j1}\lambda^{'*}_{2j1}|$. Others are the same as in
Fig.1 }
\end{figure}
\begin{figure} \label{RPVeta}
{\includegraphics[width=15cm] {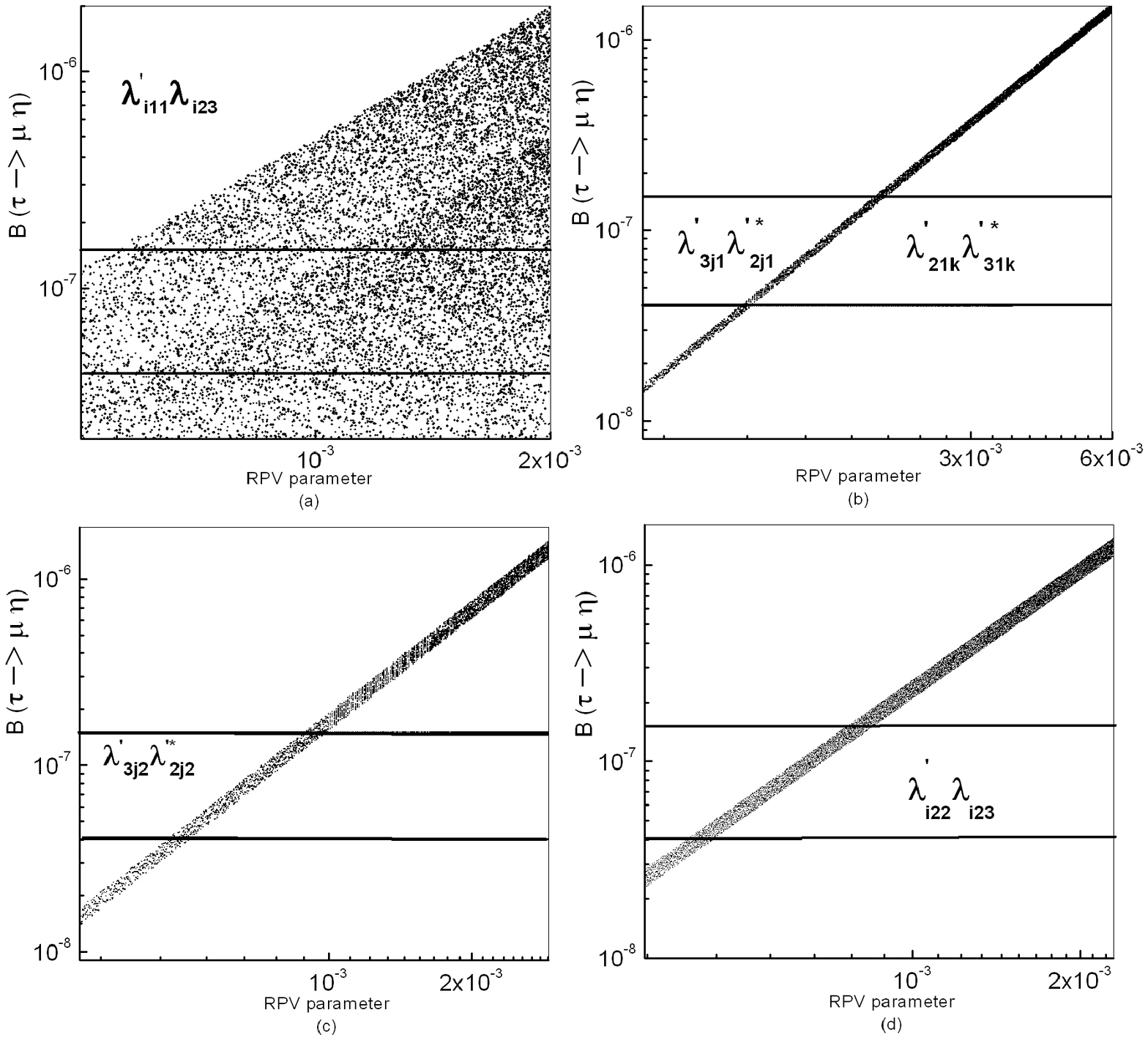}} \caption{
$\mathcal{B}(\tau\to\mu\eta)$ as function of R-parity violation
couplings.  (a) for $|\lambda'_{i11}\lambda_{i23}|$, (b)  for
$|\lambda'_{21k}\lambda^{'*}_{31k}|$ and
 $|\lambda'_{3j1}\lambda^{'*}_{2j1}|$,
 (c)  for $|\lambda'_{3j2}\lambda^{'*}_{2j2}|$, and (d)
for $|\lambda'_{i22}\lambda_{i23}|$, respectively. Others are the
same as in Fig.1 }
\end{figure}
\begin{figure} \label{RPVetap}
\scalebox{0.9}{\includegraphics[width=15cm] {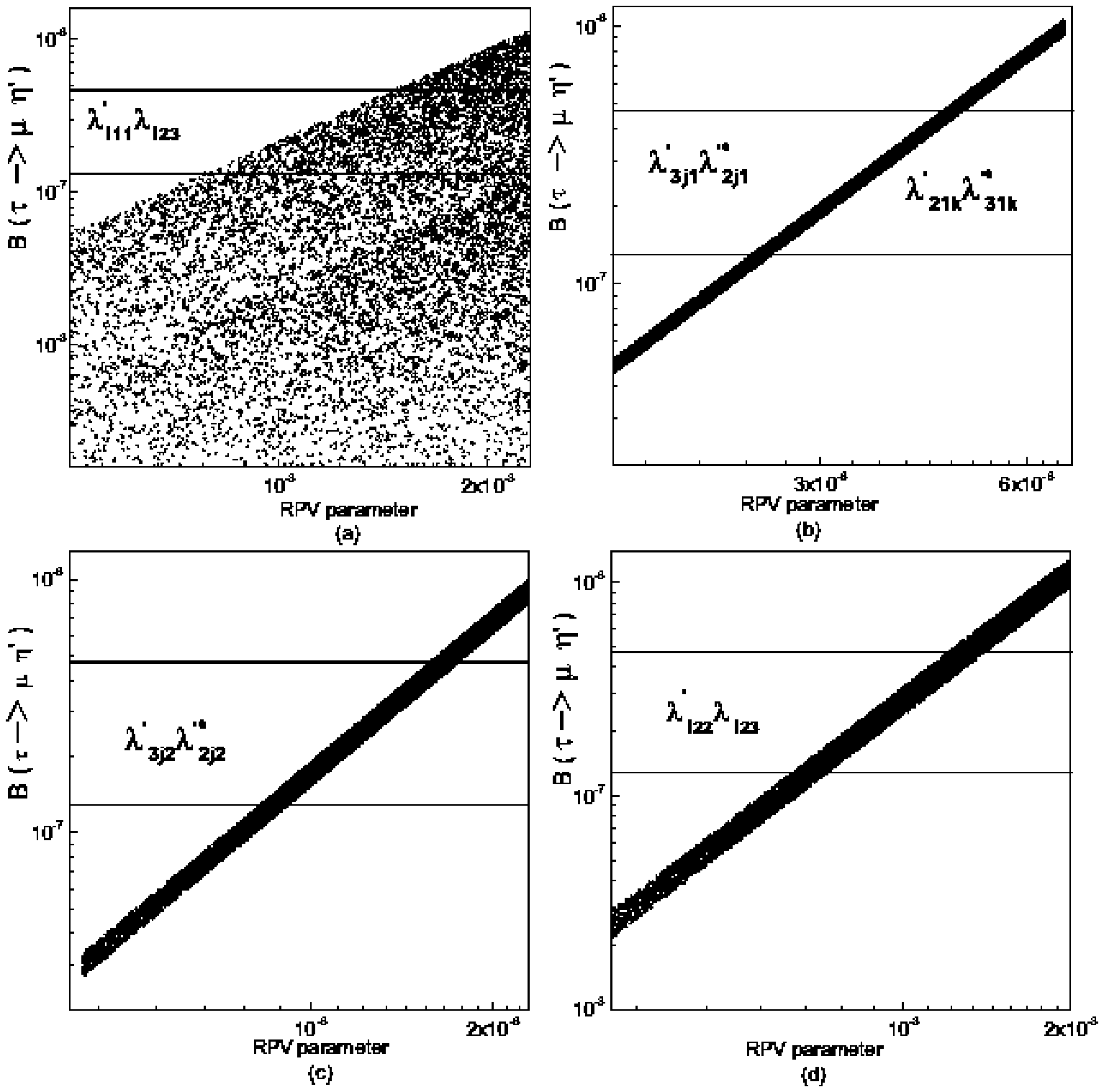}}
 \caption{ $\mathcal{B}(\tau\to\mu\eta')$ as function of R-parity
violation couplings. (a)  for $|\lambda'_{i11}\lambda_{i23}|$, (b)
for $|\lambda'_{21k}\lambda^{'*}_{31k}|$ and
 $|\lambda'_{3j1}\lambda^{'*}_{2j1}| $, (c)  for $|\lambda'_{3j2}\lambda^{'*}_{2j2}|$
, and (d) for $|\lambda'_{i22}\lambda_{i23}|$, respectively.
Others are the same as in Fig.1 }
\end{figure}

\begin{table}[b]\label{m2}
\caption{Constraints on the products of $\lambda$ and $\lambda'$
R-parity violating couplings for sfermion masses $m_{\tilde{f}} =
100 GeV$.}
\begin{center}
\begin{tabular}{c c c c c}
\hline\hline Couplings~~&~~~~$\tau \to \mu \pi^0$~~~~~~~&~~~$\tau \to \mu \eta$~~~~~~
&~~~~~~$\tau \to \mu \eta'$&~~~Previous Bounds\\
\hline
 $|\lambda'_{i11}\lambda_{i23}|$ &$\leq 1.46\times 10^{-3}$
 & &~~
 &~~~~~~$\leq 3.6\times 10^{-2}$\cite{pko},
~~$\leq1.1\times10^{-2}$\ ~\cite{Hagiwara}\\

$|\lambda'_{21k}\lambda^{'*}_{31k}|$&$\leq 1.80\times 10^{-3}$
& $\leq 1.93\times 10^{-3}$&$\leq 4.69 \times 10^{-3}$
&$|\lambda'_{i11}\lambda^{'}_{j11}|\leq 1.2\times 10^{-2}$~\cite{Barbier}\\

$|\lambda'_{3j1}\lambda^{'*}_{2j1}|$&$\leq 1.80\times 10^{-3}$
 &$\leq 1.93\times 10^{-3}$ &$\leq 4.69 \times 10^{-3}$&
$\lambda'_{3jk}\lambda^{'*}_{2jk}\leq 9.1 \times 10^{-2}$\ ~\cite{chemtob} \\

$|\lambda'_{3j2}\lambda^{'*}_{2j2}|$&&$\leq 9.38 \times 10^{-4}$
 &$\leq 1.53 \times 10^{-3}$
&$|\lambda'_{3jk}\lambda^{'*}_{ijk}|\leq 1.2\times 10^{-2}$~\cite{Barbier}\\

$|\lambda'_{i22}\lambda_{i23}|$&&$\leq 8.03\times 10^{-4}$& $\leq
1.3\times 10^{-3}$ &~~~~~~~~$\leq4.5\times 10^{-2}$\ ~\cite{pko},
$\leq 3.0\times10^{-2}$\ ~\cite{Hagiwara}\\
\hline \hline\end{tabular}
\end{center}
\end{table}
%%%%%%FIG2%%%%%%
From Eq.(\ref{rpv:mp}-\ref{rpv:mep}), we can see that the
couplings relevant to
 lepton-number violation are $\lambda$ and $\lambda'$ couplings.
Therefore, in the decays $\tau \to \mu M$ the parameters are the
products of them: $\lambda'_{21k}\lambda^{'*}_{31k},
\lambda'_{i11} \lambda_{i23},
\lambda'_{3j1}\lambda^{'*}_{2j1},\lambda'_{3j2}\lambda^{'*}_{2j2}$,
and $ \lambda'_{i22} \lambda_{i23}$ which are denoted by
$L_i(i=1,2,3,4,5)$, respectively. Among these coupling products,
the first three  contribute to $\tau \to \mu \pi^0$ decay and all
of them contribute to $\tau \to \mu \eta(\eta')$ decays. Besides
the type of $(\bar{\mu}\tau)_{S-P}$ operators,
 $(\bar{\mu}\tau)_{V\pm A}$ operators also appear in the amplitudes of these decays due
 to Fierz re-arrangements.
 Because of $\langle P^0|\bar{q}\gamma_\mu q|0 \rangle =0$,
 $\lambda'_{21k}\lambda^{'*}_{31k}$
 and $\lambda'_{3j1}\lambda^{'*}_{2j1}$
will contribute to these decays with the same coefficient.

As in literature,  we assume that only one sfermion contributes
one time with universal mass 100GeV.  We present our results in
Fig.2-4 for $\tau^- \to \mu^- \pi^0$,
 $\tau^- \to \mu^- \eta$ and $\tau^- \to \mu^- \eta'$ decays,
respectively.

From Fig.2-4, we find
 that these decays could be enhanced to the present Belle sensitivities with the
presences of RPV couplings constrained by other
processes\cite{report}, thus we can derive tighter bounds on the
relevant RPV couplings. In Table II, we present  bounds on RPV
couplings derived from the Belle upper limits on the three $\tau$
LFV decay modes\cite{0503041ex} at $90\%$ CL with central values
defaulted for hadronic inputs. Most of them are stronger than
before.

For $\tau \to \mu \pi^0$ decay, only three coupling products
$|\lambda'_{i11}\lambda_{i23}|$,
$|\lambda'_{21k}\lambda^{'*}_{31k}|$ and
 $|\lambda'_{3j1}\lambda^{'*}_{2j1}|$
contribute to branching ratios. As shown by Fig.2, the upper
bounds for $|\lambda'_{i11}\lambda_{i23}|$,
$|\lambda'_{21k}\lambda^{'*}_{31k}|$ and
 $|\lambda'_{3j1}\lambda^{'*}_{2j1}|$ are
$\mathcal{O}(10^{-3})$ which are more stringent than previous
bounds ($\mathcal{O}(10^{-2})$)\cite{Barbier,chemtob}.  It is
noted that the contribution of $|\lambda'_{i11}\lambda_{i23}|$ is
enhanced by a factor $1/m_{d}$. In numerical calculation, we take
$m_d =(4.2\pm 1.0)$MeV \cite{Hagiwara}, as shown by Fig.2(a),
which causes large uncertainties for our theoretical prediction of
$|\lambda'_{i11}\lambda_{i23}|$ contribution to $\tau\to\mu\pi^0$.
With defaulted value $m_d =4.2$MeV, we get
$|\lambda'_{i11}\lambda_{i23}|\leq 1.46\times 10^{-3}$.

All the five PRV coupling products contributing to $\tau \to \mu
\eta,\mu\eta'$ decays. The sensitivities of the decays $\tau\to
\mu\eta$ and $\tau\to \mu\eta'$ to these five RPV coupling
products are depicted by Fig.3 and Fig.4, respectively. As shown
by Fig.3.(a) and Fig.4.(a), the contributions of
$|\lambda'_{i11}\lambda_{i23}|$($L_1$) are subjected to huge
theoretical uncertainties which arise from the poorly known
$h^q_{\eta(\eta')}$ and $m_q$ as shown by Eq. (\ref{rpv:me},
\ref{rpv:mep}). We could not get meaningful upper bounds on
$|\lambda'_{i11}\lambda_{i23}|$ from $\tau \to \mu \eta,\mu\eta'$
decays.

As shown by Fig.3.b-d  and Fig.4.b-d,   $\tau \to \mu
\eta,\mu\eta'$ decays are very sensitive to  the contributions of
$|\lambda'_{21k}\lambda^{'*}_{31k}|$,
$|\lambda'_{3j1}\lambda^{'*}_{2j1}|$,
$|\lambda'_{3j2}\lambda^{'*}_{2j2}|$ and
$|\lambda'_{i22}\lambda^{'*}_{i23}|$. Therefor we get  strong
bounds on these four products which have improved the existing
ones by one order \cite{Barbier,chemtob, pko,chemtob}. From
Fig.3.b-d  and Fig.4.b-d, we can see that theoretical
uncertainties are small for predicting the contributions of these
four RPV coupling products.

It is noted that our study of $\tau\to\mu\eta' $ in RPV SUSY is
new. The decays $\tau \to \mu \pi^0, \mu\eta $ have been studied
by Kim, Ko and Lee\cite{pko}, however, we have used up-to-date
hadronic inputs for $\eta$ and $\eta'$~\cite{9812269}.

\subsection{$\tau\to \mu M$ decays in $Z'$ model with family non-universal couplings}

Many extensions of the standard model, especially grand unified
theories and supersymmetry models, have additional $Z'$ bosons. In
models with an extra $U(1)'$ gauge boson, the $Z'$  family
non-universal couplings with the SM fermions generally induce
flavor-changing neutral currents. In this paper, we refer to the
basic formalism of  $Z'$ model elaborated in Ref.\cite{lang}.

In the gauge basis, the $Z'$  neutral-current Lagrangian can be
written as
 \beq {\cal L}^{Z'} =-g'J'_\mu Z'_\mu,
 \eeq
 with $g'$the gauge coupling constant of the $U(1)'$ group at the $M_W$
scale. Here the renormalization group running between the $M_W$
and $M_{Z'}$ scales is neglected considering the uncertainties of
parameters. We assume that the $Z'$ boson has no mixing with the
SM \textit{Z} boson~\cite{0310073}. The $Z'$ current can be
written as
 \beq {\cal J}'_\mu =\sum_{i,j}\bar{\phi}^I_i
\gamma_\mu \left[
(\epsilon_{\phi_L})_{ij}P_L+(\epsilon_{\phi_R})_{ij}P_R
\right]\phi^I_j,
 \eeq
where $I$ denotes the gauge interaction eigenstates and $\epsilon_{\phi_L(R)}$
refers to the left(right)-handed chiral coupling matrix.

The fermion Yukawa matrices $Y_\phi$ in the weak eigenstate basis
can be diagonalized by unitary matrices $V_{\phi_{L,R}}$
 \be Y^{diag}_\phi =V_{\phi_{R}} Y_\phi
V^\dagger_{\phi_{L}}. \ee
 $V_{\phi_{L,R}}$ could transform $\phi^I$ into mass eigenstate fields
$\phi_{L,R}=V_{\phi_{L,R}}P_{L,R}\phi^I$. The CKM matrix is given
by the combination
 \be V_{CKM}=V_{u,L}V^\dagger_{d,L}
 \ee

Flavor changing effects(FCNC) will present  when
$\epsilon_{\phi_L(R)}$ are nondiagonal matrices. The chiral $Z'$
coupling matrices in the physical basis of fermions thus read
 \be
B^X_u=V_{uX}\epsilon_{uX}V^\dagger_{uX},\,\, B^X_d=V_{dX}
\epsilon_{dX}V^\dagger_{dX},(X=L,R)
 \ee
 where $B^X_{u(d)}$ are hermitian.

 The low energy effective Hamiltonian of $\tau \to \mu M$ decays medicated by
 $Z'$ is
 \beq {\cal H}^{Z'}_{eff}&=
 &-\frac{4G_F}{\sqrt{2}}\left( \frac{g'M_Z}{g_1M_{Z'}}\right)^2
 \left[ B^{L}_{\tau\mu}
(\bar{\mu}\gamma_\mu P_L\tau)+ B^{R}_{\tau\mu}(\bar{\mu}\gamma_\mu
P_R\tau) \right] \sum_{q}\left[ B^L_{qq}(\bar{q}\gamma_\mu P_L
q)\right]+{\it h.c}, \eeq where
$g_1=e/(\sin\theta_W\cos\theta_W)$. The diagonal elements of
$B^{L,R}$ are real for hermiticity of the effective Hamiltonian,
but the off-diagonal elements may contain weak phases. We
introduce new weak phases $\phi^{L,R}$ for $B^{L,R}_{\tau\mu}$
($B^{L,R}_{\tau\mu}=|B^{L,R}_{\tau\mu} |e^{i\phi^{L,R}}$) under
assumption of neglecting $B^R_{qq}$.
\begin{table}[htb]\label{m3}
\caption{Constraints on the values of $\xi$ in the flavor changing
$Z'$ model with family nonuniversal couplings }
\begin{tabular}{ c c c c c }
\hline \hline
 Parameters&$\tau \to \mu \pi^0$&$\tau \to \mu \eta$
 & $\tau \to \mu \eta'$&Previous Bounds\\

\hline $|\xi^{LL}|$&~~$\leq 1.81\times 10^{-3}$
 &~~$\leq 1.91\times 10^{-3}$ &~~$\leq 6.91\times
10^{-3}$&~~$\leq 0.02([-70^\circ,-55^\circ]),~\leq 0.005([-80^\circ,-30^\circ])
$\ ~\cite{0310073},\\
& & & &~~ $\leq 0.0055(110^\circ),~\leq 0.0098(-97^\circ)$\ ~\cite{0406126} \\
 $|\xi^{RL}|$&~~$\leq 1.81\times 10^{-3}$
 &~~$\leq 1.91\times 10^{-3}$ &~~$\leq 6.91\times 10^{-3}$&\\

 $|\xi^{LL}=\xi^{RL}|$&~~$\leq 1.28\times 10^{-3}$
 &~~$\leq 1.34 \times 10^{-3}$&~~$\leq 4.78\times
 10^{-3}$&~~$\leq 0.02([5^\circ,15^\circ]),~~
 \leq0.005([-80^\circ,-30^\circ])$\ ~\cite{0310073}\\

 & & & & ~~$\leq0.0104(-70^\circ),~\leq0.0186(83^\circ)
 $~\cite{0406126}\\
\hline \hline\end{tabular}
\end{table}
Now we can write out the decay amplitudes
\begin{eqnarray}\label{dmez1}
{\cal A}(\tau^- \to \mu^- \pi^0)&=& -i \frac{G_F}{2} f_{\pi^0} p^{\mu}_{\pi^0}
 \left\{X(\bar{\mu}\tau)_{V-A}+
Y(\bar{\mu}\tau)_{V+A}
  \right\}, \\
\label{dmez2}
{\cal A}(\tau^- \to \mu^- \eta)&=& -i\frac{G_F}{\sqrt{2}}p^{\mu}_\eta
 \left\{\left[\frac{1}{2} f^q_{\eta}
 \Delta_1 + f^s_{\eta} \Delta_2\right ](\bar{\mu}\tau)_{V-A}+
\left[\frac{1}{2} f^q_{\eta}\Gamma_1+ f^s_{\eta} \Gamma_2\right]
(\bar{\mu}\tau)_{V+A}
  \right\}, \\
\label{dmez3}{\cal A}(\tau^- \to \mu^- \eta')&=&-i\frac{G_F}{\sqrt{2}}
 p^{\mu}_{\eta'}\left\{\left[\frac{1}{2} f^q_{\eta'}
 \Delta_1 + f^s_{\eta'} \Delta_2 \right ]
 (\bar{\mu}\tau)_{V-A}+
\left[\frac{1}{2} f^q_{\eta'}  \Gamma_1 + f^s_{\eta'}
\Gamma_2\right] (\bar{\mu}\tau)_{V+A}
  \right\},\end{eqnarray} with
  \begin{eqnarray}
X&=&\biggl(\frac{g'M_Z}{g_1M_{Z'}}\biggl)^2\biggl(B^{L}_{\tau\mu}
B^L_{uu}-B^{L}_{\tau\mu}B^L_{dd}\biggl),\ \ \
Y=\biggl(\frac{g'M_Z}{g_1M_{Z'}}\biggl)^2\biggl( B^{R}_{\tau\mu}B^L_{uu}-B^{R}_{\tau\mu}
B^L_{dd}\biggl), \\
\Delta_1&=&\biggl(\frac{g'M_Z}{g_1M_{Z'}}\biggl)^2\biggl(B^{L}_{\tau\mu}
B^L_{uu}+B^{L}_{\tau\mu}B^L_{dd}\biggl),\ \ \
\Delta_2=\biggl(\frac{g'M_Z}{g_1M_{Z'}}\biggl)^2B^{L}_{\tau\mu}B^L_{ss}, \ \ \\
\Gamma_1&=&\biggl(\frac{g'M_Z}{g_1M_{Z'}}\biggl)^2\biggl(B^{R}_{\tau\mu}
B^L_{uu}+B^{R}_{\tau\mu}B^L_{dd}\biggl),\ \ \
\Gamma_2=\biggl(\frac{g'M_Z}{g_1M_{Z'}}\biggl
 )^2B^{R}_{\tau\mu}B^L_{ss}.
\end{eqnarray}

%%%%%%%%FIG3%%%%%%%%%%%%%%%%%%%%%%
\begin{figure*}\label{Zpi}
\scalebox{1.0}{\includegraphics[width=15cm]{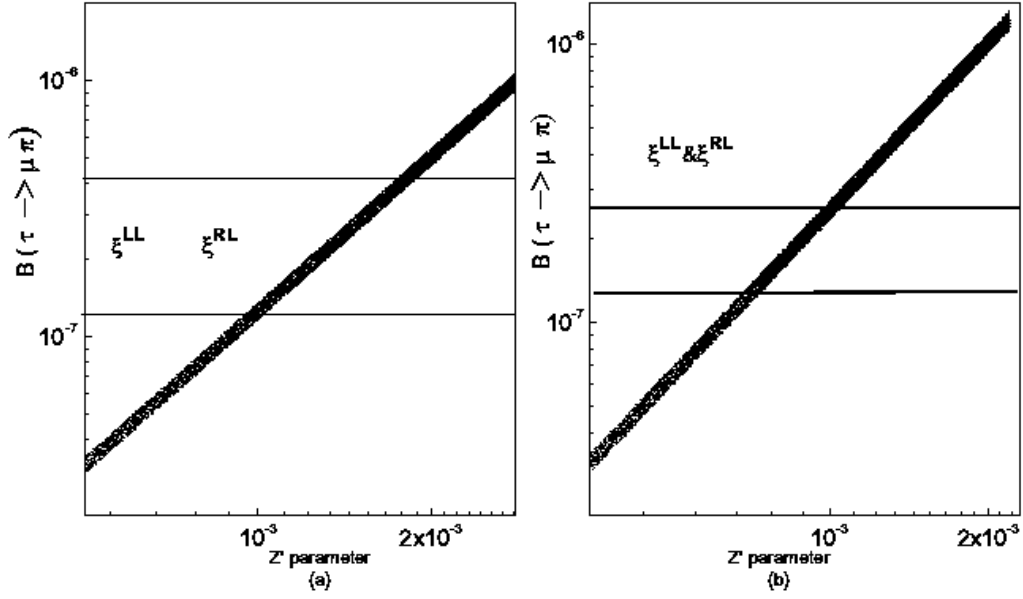}} \caption{
$\mathcal{B}(\tau\to \mu \pi^{0})$ as function of $Z'$ couplings.
(a) for $|\xi^{LL}|$ and  $|\xi^{RL}|$ contributing seperatly.
 (b)  for both
$|\xi^{LL}|$ and $|\xi^{RL}|$ contributing with the same strength.
 The horizontal lines are the same as these in Fig.1.}
\end{figure*}
\begin{figure}\label{Zeta}
\scalebox{1.6}{\includegraphics[width=8cm]{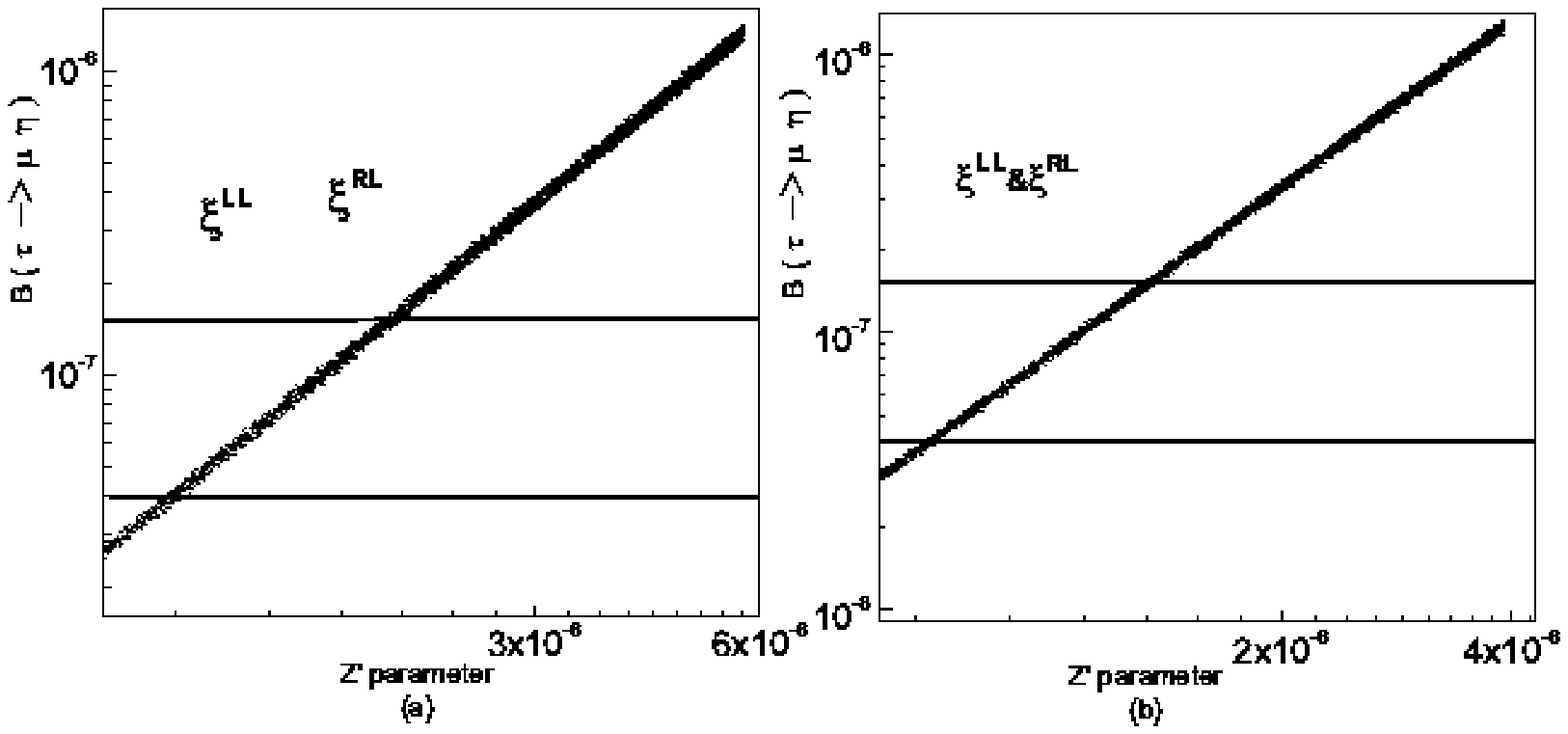}} \caption{
$\mathcal{B}(\tau\to \mu \eta)$ as function of $Z'$ couplings. The
horizontal lines are same as these in Fig.1.}
\end{figure}
\begin{figure*}\label{Zetap}
\scalebox{0.9}{\includegraphics[width=15cm]{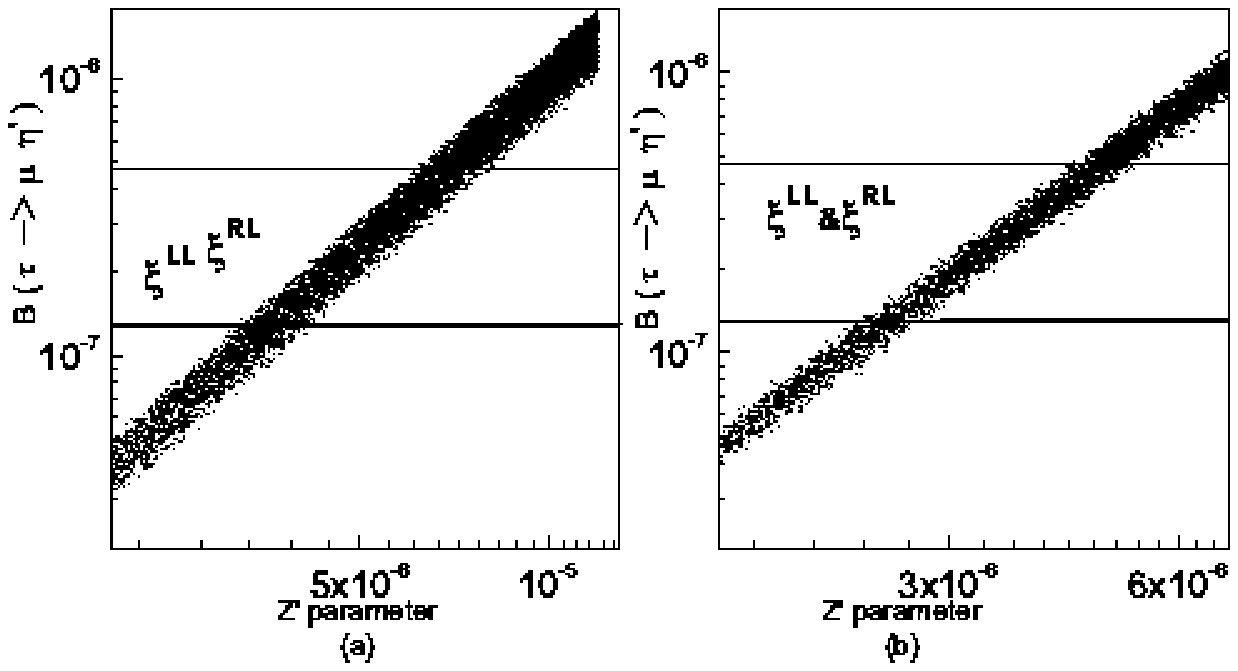}} \caption{
$\mathcal{B}(\tau\to \mu \eta')$ as function of $Z'$ couplings.
The horizontal lines are same as these in Fig.1.}
\end{figure*}
%%%%%%%%%%%%%%%%%%%%%%%%%%%%%%%

 As in literature, we take~\cite{0310073}
\beq
\xi^{LL}=(\frac{g'M_Z}{g_1M_{Z'}})^2 B^{L}_{\tau\mu} B^L_{uu}e^{i\phi_L}, \ \ \
 \xi^{RL}=(\frac{g'M_Z}{g_1M_{Z'}})^2 B^{R}_{\tau\mu} B^L_{uu}e^{i\phi_R},
 \eeq
and $B^L_{uu}\simeq -2 B^L_{dd(ss)}$~\cite{0406126}.

 We find that the variation of phase $\phi_{L,R}$ have nearly
 negligible influence on the
branching ratios. So we take $\phi_{L,R}=10^\circ$~\cite{0310073} as benchmark.
At first, we assume that only one parameter contributes one time and
then both
 $\xi^{LL}$ and $\xi^{RR}$ contribute
with the same strength. We present the correlation of branching
ratio versus parameter $|\xi|$ in Fig.5-7.   The branching ratios
 get the same contributions from $|\xi^{LL}|$ and $|\xi^{RL}|$.
 The upper bounds on parameters $|\xi|$ from experimental data
 are $\mathcal{O}(10^{-3})$ which are listed in Table III for comparison.
 Compared to former two new physics scenarios, there is no scalar operator induced
 by  $Z'$ family non-universal couplings. Thus the poorly known hadronic parameters
  $h^{q,s}_{M}$ and factor $1/m_{q}$ are absent. Theoretical uncertainties
are  due to decay constants $f_{\pi}$ and $f^{q,s}_{\eta,\eta'}$
which have been determined with few percentages accuracy  as
listed in Eq.(10). So theoretical estimations of these decays
could be made quite accurate.

As shown by Fig.5 and 6, the sensitivities of
$\mathcal{B}(\tau\to\mu\pi^0)$ and $\mathcal{B}(\tau\to\mu\pi^0)$
to the $Z'$ LFV FCNC couplings are quite similar. From Belle upper
limit on $\mathcal{B}(\tau\to\mu\pi^0)$,  we get $\xi < 2\times
10^{-3} $ which are comparable with previous bounds
\cite{0310073,0406126}. However,
 from Fig.7, we can see that  $\tau \to \mu \eta'$ decay
gives weaker bounds  than $\tau \to \mu \eta(\pi)$.

Anyhow the Belle searching for  $\tau \to \mu M$ decays have
already put strong bounds on the parameter spaces. In other words,
 these decays could be enhanced to the present B factories sensitives
 by the $Z'$ family nonuniversal FCNC couplings without conflict with bounds
 from  other exist measurements.

\section{Conclusion}

The measurement of LFV processes would be a definite evidence for
physics beyond the SM. In this paper, we have studied  LFV
processes $\tau^- \to \mu^- M (M=\pi^0,\eta,\eta')$ at the tree
level in the 2HDM III, RPV SUSY and flavor changing $Z'$ model
with family non-universal couplings. Since these decays are very
sensitive to the presences of LFV couplings, we have derived
constraints on parameter space of the three New Physics scenarios
from the recent Belle limits\cite{0503041ex}. Our main findings
can be summarized as follows.

\begin{enumerate}
\item  In 2HDM III, the strongest bound on $\lambda_{\tau\mu}$
comes from $\tau\to\mu\eta$ decay. The bound is consistent with
these in literature, however, improve these by several times.

 \item  In RPV SUSY,  $\tau \to \mu \eta,\mu\eta'$ decays are very
sensitive to the contributions of
$|\lambda'_{21k}\lambda^{'*}_{31k}|$,
$|\lambda'_{3j1}\lambda^{'*}_{2j1}|$,
$|\lambda'_{3j2}\lambda^{'*}_{2j2}|$ and
$|\lambda'_{i22}\lambda^{'*}_{i23}|$
 Therefor we get
strong bounds on these three products which have improved the
existing ones by one  order \cite{Barbier,chemtob, pko,chemtob}.
However, there are large  uncertainties in calculating
contributions of these RPV coupling products due to hadronic
parameters $h^{s}_{\eta,\eta'}$.  We could not get bounds on
$|\lambda'_{i11}\lambda^{'*}_{i23}|$ from the two decays because
of poorly known   $h^{s}_{\eta,\eta'}$. However, we can get strong
bound on $|\lambda'_{i11}\lambda^{'*}_{i23}|$ with $\tau\to
\mu\pi^0$.

\item  In $Z'$ model, theoretical predictions of $Z'$ LFV coupling
contributions could be make quite accurate.  $\tau \to \mu \pi,
\mu \eta$ and $\mu\eta'$ decays have similar sensitivities  to the
LFV couplings. Belle current upper limits on $B(\tau \to \mu
\eta,\mu\eta' )$ have already constrained $\xi$ as small as
$\mathcal{O}(10^{-3})$.
\end{enumerate}

In summary, we have shown that the LFV semileptonic  decays $\tau
\to \mu \pi^0,\,
 \mu\eta$ and $\mu\eta'$ are very sensitive to the presences of LFV
couplings in 2HDM III, RPV SUSY and $Z'$ models. Using
\textit{only} $153.8fb^{-1}$ data, Belle recent upper limits  for
these decays have already given quite tight bounds on the LFV
couplings in the aforementioned three new physics scenarios. It
should be noted that Belle and BaBar had accumulated about
$540fb^{-1}$ and $320fb^{-1}$, respectively, till the end of year
2005. With refined measurements with the these data at Belle and
BaBar, we could  get more crucial information on LFV, at least
more stringent bounds on the parameter spaces of models with LFV
couplings. The results derived in this paper from the recent
measurements at Belle would be useful for phenomenological studies
of the scenarios in other interesting processes.

\begin{acknowledgments}
The work is supported  by National Science Foundation under contract No.10305003,
Henan Provincial Foundation for Prominent Young Scientists under
contract No.0312001700 and the NCET Program sponsored by Ministry of Education, China.
\end{acknowledgments}

\end{document}